# Liquid Neural Networks: Next-Generation AI for Telecom from First Principles

ZHU Fenghao[1], WANG Xinquan[1], ZHU Chen[2], HUANG Chongwen[1]

(1. Zhejiang University, Hangzhou 310027, China;
2. Polytechnic Institute, Zhejiang University, Hangzhou 310015, China)

**Abstract:** Artificial intelligence (AI) has emerged as a transformative technology with immense potential to reshape the next-generation of wireless networks. By leveraging advanced algorithms and machine learning techniques, AI offers unprecedented capabilities in optimizing network performance, enhancing data processing efficiency, and enabling smarter decision-making processes. However, existing AI solutions face significant challenges in terms of robustness and interpretability. Specifically, current AI models exhibit substantial performance degradation in dynamic environments with varying data distributions, and the black-box nature of these algorithms raises concerns regarding safety, transparency, and fairness. This presents a major challenge in integrating AI into practical communication systems. Recently, a novel type of neural network, known as the liquid neural networks (LNNs), has been designed from first principles to address these issues. In this paper, we explore the potential of LNNs in telecommunications. First, we illustrate the mechanisms of LNNs and highlight their unique advantages over traditional networks. Then we unveil the opportunities that LNNs bring to future wireless networks. Furthermore, we discuss the challenges and design directions for the implementation of LNNs. Finally, we summarize the performance of LNNs in two case studies.

## 1. Introduction

The sixth-generation (6G) wireless communication network is envisioned to revolutionize the telecommunications landscape by incorporating a wide range of advanced communication capabilities. These enhancements are expected to support enriched and immersive experiences, ensure ubiquitous and seamless coverage, and enable innovative forms of collaboration [1]. One of the primary catalysts for these advancements is the incorporation of technologies driven by artificial intelligence (AI). By harnessing AI, 6G aims to overcome the limitations of current networks, offering ultra-reliable and low-latency communication, extensive connectivity for massive IoT devices, and significantly improved mobile broadband services [2]. This integration will facilitate ground-breaking applications such as holographic telepresence, tactile internet, intelligent autonomous systems, and smart cities. By leveraging AI, 6G networks will not only optimize performance and enhance data processing efficiency but also ensure adaptive, secure, and robust communication environments [3-5]. The synergy between AI and 6G will pave the way for unprecedented connectivity and intelligent communication solutions, fundamentally transforming industries and society as a whole.

Although AI has exhibited great potential in reshaping the next generation of wireless networks,

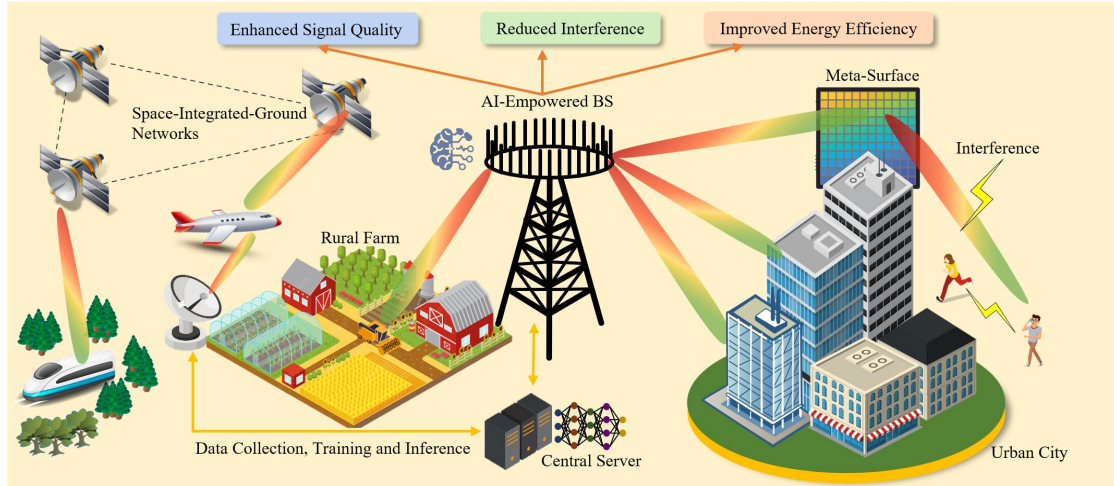

Fig. 1. 6G communication scenarios with AI integration.

deploying it in practical communication scenarios remains challenging [6-8]. Fig. 1 shows the 6G communication scenarios with AI integration. These challenges arise primarily due to several factors, as outlined below:

1) **The Issue of Robustness**: Current AI models often struggle to maintain performance in dynamic and unpredictable environments with varying data distributions [9]. In practical wireless networks, conditions such as user mobility, changing signal conditions, and interference can cause significant variations in the data fed into AI models. This lack of robustness can lead to substantial performance degradation, making it difficult to ensure reliable operation [6]. For instance, an AI model trained under certain static conditions may fail to adapt when deployed in a real-world setting where wireless parameters continuously change. Addressing this issue requires the development of adaptive AI models that can learn and generalize from a wide range of conditions and data patterns.

2) **The Issue of Interpretability**: The black-box nature of many AI algorithms poses significant challenges in understanding and explaining their decision-making processes [7]. This lack of interpretability raises concerns about the safety, transparency, and fairness of AI-driven solutions in communication systems. In wireless networks, where decisions can impact a wide range of users and services, it is crucial to ensure that AI models make decisions that are understandable and justifiable. For example, in the context of spectrum resource allocation in communications, the AI system should make fair and efficient decisions. Enhancing the interpretability of AI models involves developing methods to provide insights into the inner workings and decision criteria of the models.

3) **The Issue of Complexity**: AI models, particularly deep learning architectures, require substantial computational resources and complex infrastructure for training and deployment. Training deep neural networks can be computationally intensive and time-consuming, often necessitating specialized hardware such as GPUs or TPUs. Moreover, the deployment of these models in practical wireless networks must contend with constraints such as limited bandwidth, low latency requirements, and the need for real-time processing [8]. This high complexity can hinder their practical implementation, especially in edge computing scenarios where computational resources are limited. Simplifying AI models and optimizing their performance to run efficiently on resource-constrained devices are critical areas of research.

Addressing these challenges is crucial for the successful integration of AI into next-generation wireless networks, paving the way for more reliable, transparent, and efficient communication systems. Overcoming these hurdles will allow AI to revolutionize wireless communication, enhancing user experiences, improving network performance, and enabling innovative applications and services. In this article, we provide an overview of the recently proposed liquid neural networks (LNNs) [10-12], which are designed from first principles to be robust, interpretable, and resource-efficient, making them well-suited for the dynamic and complex nature of wireless communication environments. We explore the opportunities that LNNs bring to future wireless networks and discuss the challenges and design directions for their implementation.

The rest of this paper is organized as follows: Section 2 provides an overview of traditional neural network and their limitations. Section 3 describes the design of LNNs. Section 4 describes the features and benefits of LNNs. Section 5 explores the opportunities that LNNs bring to the future wireless networks. Section 6 discusses the main challenges associated with LNN-based communication systems and outlines potential future research directions. Case studies are presented in Section 7 to verify the performance of LNNs. Finally, the conclusions are summarized in Section 8.

## 2. Overview of Traditional Neural Networks

Traditional neural networks are fundamental in the development of AI technologies, each offering unique strengths for various applications. Below, we discuss four primary types of traditional neural networks: feed forward neural networks (FNNs), convolutional neural networks (CNNs), recurrent neural networks (RNNs), and ordinary differential equation neural networks (ODE-NNs).

2.1 Feed Forward Neural Networks
FNNs are the simplest type of artificial neural network architecture. In FNNs, the information moves in one direction—from input nodes, through hidden nodes (if any), to output nodes [13]. There are no cycles or loops in the network. This structure makes FNNs suitable for simple pattern recognition tasks, such as image classification or function approximation. However, they may struggle with tasks requiring memory or temporal dependencies due to their lack of internal hidden states.

2.2 Convolutional Neural Networks
CNNs are specialized for processing structured grid data, such as images. They utilize convolutional layers that apply filters to the input data to capture spatial hierarchies of features [14]. CNNs are highly effective in image and video recognition, owing to their ability to learn spatial hierarchies and patterns. The architecture typically includes convolutional layers, pooling layers, and fully connected layers. CNNs are known for their robustness in handling variations in image data, such as shifts, scales, and distortions.

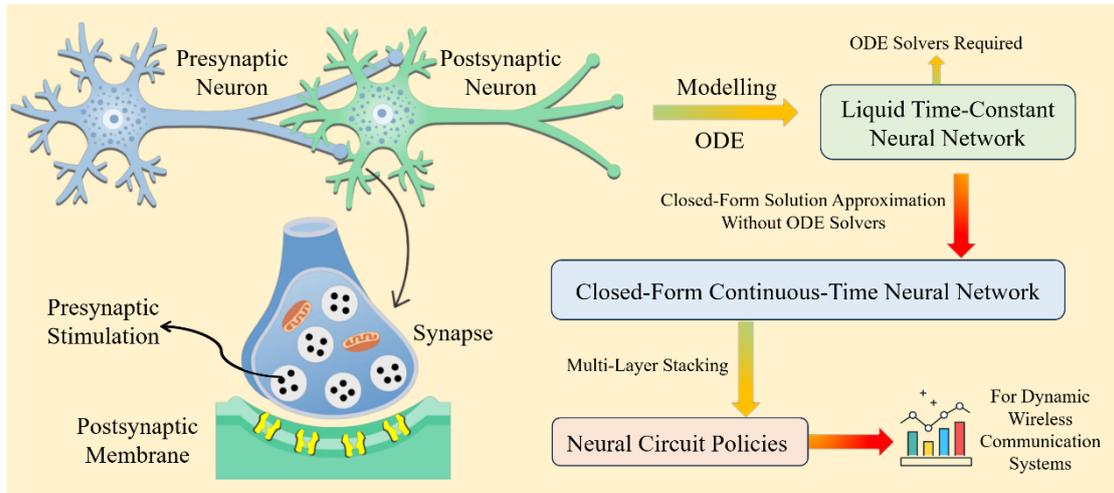

Fig. 2. Liquid neuron and the ODE modelling.

2.3 Recurrent Neural Networks

RNNs are designed to handle sequential data and capture temporal dependencies. Unlike FNNs, RNNs have connections that form directed cycles, allowing information to persist across different steps in the sequence. This makes RNNs powerful for tasks like time series data, natural language processing, and speech recognition [15]. However, traditional RNNs can suffer from vanishing and exploding gradients, which hinder their performance on long sequences. To address these issues, variants such as long short-term memory (LSTM) networks and gated recurrent units (GRUs) were developed. LSTMs introduce gating mechanisms to manage long-term dependencies, while GRUs simplify the architecture for computational efficiency. Despite these improvements, both LSTMs and GRUs have limitations, including the inability to model continuous-time dynamics and reduced robustness in highly dynamic environments [16].

2.4 Ordinary Differential Equation Neural Networks

ODE-NNs are designed to model continuous-time dynamics, addressing limitations of traditional RNNs [17]. continuous-time recurrent neural networks (CT-RNNs) and ODE-LSTM networks are key examples. CT-RNNs use ODEs to capture continuous-time sequences, making them suitable for irregular time intervals, but they are computationally intensive due to the need for numerical solvers. ODE-LSTMs integrate continuous-time modeling into the LSTM framework, enhancing their ability to handle continuous dependencies. Despite these improvements, ODE-LSTMs and CT-RNNs face challenges such as increased computational complexity and potential training instability, which can limit their effectiveness and robustness in highly dynamic environments.

## 3. Design of LNNs

LNNs are uniquely designed based on first principles, fundamentally differing from other models in their neuron operation [10]. First principles involve deriving properties and behaviors directly from fundamental laws of nature, ensuring that the design is grounded in the most essential elements. Inspired by the dynamic and adaptive nature of biological neural systems, LNNs mimic the information transmission mechanisms observed at synapses in the nematode Caenorhabditis elegans. This approach enables LNNs to emulate the flexibility and resilience of natural neural networks. Unlike static architectures, LNNs can continuously adapt and reorganize in response to new inputs, maintaining high performance and robustness in dynamic and unpredictable environments. This adaptability makes LNNs particularly well-suited for real-world applications where conditions constantly change. Currently, there are three types of liquid neural networks: liquid time-constant neural networks (LTCs), closed-form continuous-time neural networks (CfCs), and neural circuit policies (NCPs).

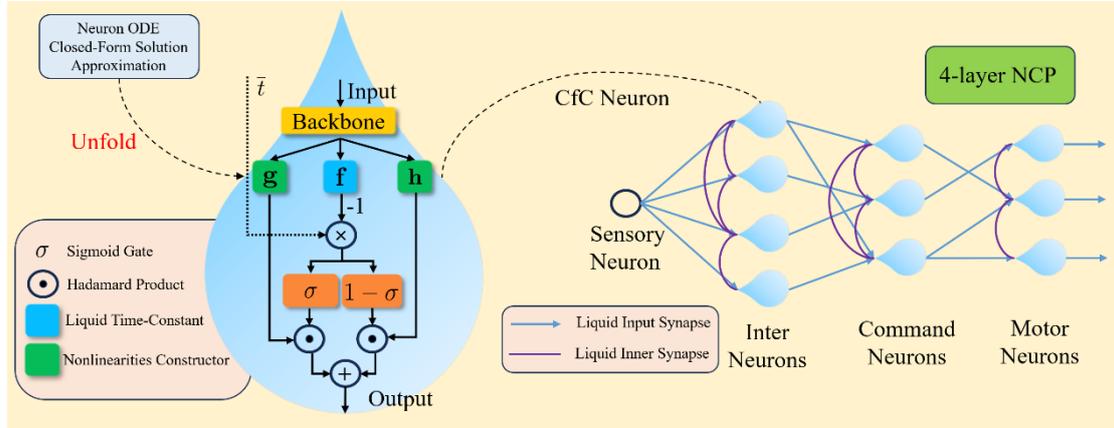

Fig. 3. The structure of a CfC neuron and a 4-layer NCP.

3.1 Liquid Time-constant Neural Networks

Fig. 2 illustrates the basic information flow of a liquid neuron, which serves as the fundamental building block of LTCs. In this model, a presynaptic neuron transmits information to a postsynaptic neuron via the synapse between them, using presynaptic stimuli. The potential of the postsynaptic membrane acts as a dynamic variable, representing the hidden states in the corresponding neural networks. This entire process is described by an ordinary differential equation (ODE), which captures the dynamic, non-linear interactions between neurons. LTCs have demonstrated exceptional flexibility and generalizability, particularly in applications such as vehicle autopilot and vehicular communications. These networks can adapt to various changing external conditions with remarkable efficiency. Notably, LTCs have achieved high-fidelity autonomy in complex autonomous systems with as few as 19 liquid neurons [11]. This capability can be extended to enhance vehicle-to-everything (V2X) communications, where LTCs can optimize data transmission and processing in dynamic, real-time environments. By integrating LTCs into next-generation wireless communication systems, vehicles can achieve seamless connectivity, improved network performance, and robust decision-making processes. This enables sophisticated or task-specific operations even under diverse and fluctuating conditions.

3.2 Closed-form Continuous-time Neural Networks

While LTCs can adapt to changing environments, their lack of a closed-form solution requires computationally intensive iterative solvers for forward propagation and backpropagation. To address this issue, a closed-form solution was proposed to approximate the true solution of the ODE [12], as illustrated in Fig. 2. The closed-form expression successfully circumvents the high overhead of traditional ODE solvers and approximates the solution with a few parameters. To take advantage of existing deep learning tools and theories, CfCs are represented by a specially designed deep neural network structure, as depicted in Fig. 3. This innovative approach significantly reduces computational complexity while maintaining the adaptability and robustness characteristic of liquid neural networks, making CfCs highly suitable for real-time applications in dynamic environments.

3.3 Neural Circuit Policies

To further exploit the potential of LTCs and CfCs, NCPs are designed to combine multiple CfC or LTC neurons into several layers. An example of an NCP comprising multiple CfC neurons is illustrated in Fig. 3. A typical NCP features four distinct layers: the sensory neuron layer, the inter neurons layer, the command neurons layer, and the motor neurons layer. These layers feature sparse connections both within and between them, mimicking the sparse connectivity observed in biological neural networks. This design reduces computational complexity and accelerates information exchange and fusion. NCPs have demonstrated robust flight navigation capabilities when presented with out-of-distribution data, generalizing effectively to scenarios that were not encountered during training [18]. This ability to handle new and diverse conditions makes NCPs highly valuable for applications requiring high adaptability and real-time processing in dynamic environments.

## 4. Features and Benefits of LNNs

LNNs stand out due to their unique design and operational principles, which endow them with several distinct features and benefits over traditional neural network models. These characteristics make LNNs exceptionally well-suited for the dynamic and complex nature of modern wireless communication systems. In this section, we delve into the key features and benefits of LNNs, highlighting their superior expressivity, generalizability, interpretability, lower complexity, and continuous-time modeling capabilities.

4.1 Superior Generalizability and Robustness

LNNs exhibit superior generalizability and robustness compared to traditional neural networks, primarily due to their biologically inspired design that allows continuous adaptation to new and varying inputs [18]. This dynamic and adaptive nature mimics biological neural systems, enabling LNNs to generalize well across different conditions and environments. Such adaptability and resilience are particularly valuable in wireless communications, where network conditions and user demands can change rapidly. LNNs maintain high performance even when faced with data that deviates significantly from the training set, making them ideal for real-world applications where unpredictability is the norm. For instance, in dynamic spectrum access and adaptive beamforming, LNNs can adjust to varying spectrum availability and signal conditions in real-time, ensuring optimal communication performance. Moreover, their ability to continuously reorganize and adapt enhances their robustness, allowing them to handle unexpected changes and disturbances effectively. This capability is crucial for maintaining reliable communication links in highly dynamic and unpredictable environments, such as during emergency response scenarios. In such cases, LNNs can adapt to fluctuating network topologies and varying signal conditions, ensuring the delivery of critical information and maintaining robust and resilient communication networks.

### 4.2 Enhanced Expressivity

LNNs exhibit enhanced expressivity compared to traditional neural networks due to their ability to dynamically adapt to incoming data and capture intricate temporal patterns. This expressivity is evident in their ability to generate complex latent space trajectories when exposed to various input patterns. For example, LNNs produce significantly more detailed and longer trajectories than models like Neural ODEs and Continuous-Time RNNs [10], indicating a higher capacity for nuanced temporal representation. This enhanced expressivity directly contributes to their ability to quickly adapt to changing conditions. The complex internal representations allow LNNs to effectively process and integrate new information, enabling rapid adjustments to new inputs and environments. In wireless communication, this means LNNs can adapt to rapidly changing network conditions and user behaviors, ensuring consistent and reliable performance. Furthermore, the continuous adaptation mechanisms of LNNs, inspired by biological neural systems, support their superior expressivity. This adaptability enables LNNs to maintain high levels of detail and accuracy in their representations, even in dynamic and unpredictable environments, making them ideal for complex and varied tasks in advanced telecommunications applications.

### 4.3 Improved Interpretability

LNNs offer significantly improved interpretability compared to traditional neural networks, an advantage particularly important for applications requiring transparency and trust. The interpretability of LNNs arises from their ability to disentangle complex neural dynamics into comprehensible and distinct behaviors. By leveraging techniques such as decision trees to analyze neural policies, LNNs can provide clear and logical explanations for their decision-making processes [7]. This enhanced interpretability is essential for understanding and debugging model behavior, especially in safety-critical systems like robotics, autonomous driving, and dynamic wireless communication networks. In the context of telecommunications, interpretability is crucial for ensuring reliable network performance and facilitating troubleshooting. For instance, understanding how LNNs manage spectrum allocation or adjust beamforming in real-time can help network operators optimize resource usage and maintain robust connectivity. Disentangling neural responses into identifiable strategies and behaviors allows for a more precise evaluation of how well these networks capture and represent underlying task dynamics in communication systems. This capability not only boosts the trustworthiness of LNNs but also facilitates their deployment in real-world scenarios where understanding the rationale behind decisions is crucial for maintaining high-performance and reliable wireless communications.

### 4.4 Lower Complexity

LNNs benefit from lower computational complexity due to their efficient design, arising from several factors. First, the sparse connectivity within and between the layers of NCPs reduces computational overhead, making the networks more efficient without compromising performance.

Second, the closed-form solutions used in LNNs eliminate the need for complex iterative solvers typically required for solving ordinary differential equations (ODEs), further lowering computational complexity. Additionally, LNNs possess strong expressive power, enabling them to perform complex tasks with fewer neurons, significantly reducing overall size of the network. This combination of factors is particularly beneficial in applications requiring real-time processing and decision-making, such as V2X communications and dynamic wireless networks. In telecommunications, lower complexity translates to faster processing speed and reduced energy consumption, which are critical for the scalability and sustainability of next-generation networks [19]. Efficient resource allocation, real-time traffic management, and rapid handovers in mobile networks all benefit from the lower complexity of LNNs, leading to more efficient and robust communication systems. Moreover, the low complexity and efficient design of LNNs contribute to energy savings and environmental sustainability. This makes LNNs suitable for deployment in environments with limited computational and energy resources, including edge devices and IoT sensors where computational power and battery life are constrained. By reducing energy consumption, LNNs support a wide range of applications, from smart cities to remote monitoring systems, without overwhelming the network infrastructure, thereby promoting greener and more sustainable technology solutions.

4.5 Continuous-Time Modelling

One of the distinctive features of LNNs is their continuous-time modeling capability. Unlike traditional neural networks that operate in discrete time steps, LNNs leverage ODEs to model the dynamic interactions between neurons. This allows LNNs to capture the continuous and fluid nature of real-world processes more accurately. Continuous-time modeling is particularly advantageous in scenarios requiring high temporal resolution and precision, such as real-time autonomous systems and adaptive communication networks [16]. In telecommunications, continuous-time modeling enables more precise channel estimation, interference management, and adaptive modulation schemes. By modeling the system dynamics in continuous time, LNNs can respond more naturally and effectively to the ever-changing conditions of the environment, ensuring optimal performance in rapidly varying communication scenarios [20].

In summary, LNNs offer distinct features and benefits that make them exceptionally suitable for next-generation wireless communication systems. Their superior generalizability, interpretability, lower complexity, continuous-time modeling, enhanced robustness, and efficient resource utilization position them as a transformative technology. By integrating LNNs into telecommunications, we can achieve more adaptive, reliable, and efficient networks that meet the demands of future wireless communication environments.

## 5. LNNs for Wireless

In this section, we unveil the opportunities that LNNs bring to the evolution and enhancement of future wireless networks. Specifically, we introduce two key topics: integrated sensing and communication (ISAC) and self-organizing networks (SONs).

5.1 Integrated Sensing and Communication

ISAC represents a paradigm shift in wireless network design, merging communication and sensing functionalities into a unified framework to enhance spectral efficiency (SE) and reduce hardware costs [21-22]. With the biologically inspired architecture and low computational complexity, LNNs are particularly well-suited for ISAC systems. Their ability to learn and adapt in real-time, handle complex and dynamic environments, and generalize effectively makes them ideal for optimizing resource allocation between communication and sensing functions, thus improving SE without extensive computational resources. The adaptability and interpretability of LNNs are crucial for applications like autonomous driving, where precise and reliable sensing is critical for safety, as they can learn from historical data and adapt to new conditions, enhancing sensing accuracy and reliability. In wireless communication, LNNs manage interference, optimize transmission parameters, and ensure robust links, maintaining high Quality of Service (QoS) even in challenging environments. Their low computational complexity leads to shorter processing times and reduced energy consumption, essential for scalable and sustainable next-generation green networks. Leveraging the low complexity and adaptability of LNNs, it is promising to develop joint sensing and communication algorithms, real-time learning and adaptation frameworks, and ensure compatibility with existing network infrastructure and protocols. These advancements can lead to more efficient, reliable, and versatile wireless networks.

5.2 Self-Organizing Networks

SONs represent a kind of wireless networks that are characterized by their ability to adapt and evolve autonomously in response to changing environmental conditions, network demands, and user behaviors. Unlike traditional static network configurations, these networks can dynamically reconfigure themselves, optimize resource allocation, and maintain robust performance without human intervention. With the continuous adaptation and learning capabilities, LNNs are uniquely suited for implementing SONs. Their biologically inspired architecture allows them to learn and adjust in real-time, providing seamless adaptability to varying network conditions. This is particularly crucial in wireless environments where factors such as signal interference, user mobility, and fluctuating demand can significantly impact network performance. For instance, in SONs, LNNs can be employed to predict and address potential network congestion in advance by reallocating resources or adjusting transmission parameters. They can also enhance QoS by dynamically adapting to the quality of the communication links and optimizing handovers in mobile networks. Moreover, LNNs can facilitate proactive maintenance of the communication systems by identifying and mitigating faults or anomalies before they escalate into significant issues. To sum up, SONs powered by LNNs are envisioned to revolutionize wireless communication by enabling networks that are not only more resilient and efficient but also capable of autonomously evolving to meet the ever-changing demands of users and applications. This represents a significant step towards the realization of truly intelligent and adaptive wireless networks.

## 6. Challenges and Future Research Directions

In this section, we present some of the main challenges associated with LNN-based communication systems and outline potential future research directions. The following subsections delve into specific areas where advancements are needed to fully realize the potential of LNNs in wireless communication.

## 6.1 Zero Shot Learning

Zero-shot learning (ZSL) describes a model's ability to recognize and categorize data from classes it has never seen before. This feature is fundamental for LNNs operating in ever-changing and uncertain wireless communication environments. Traditional machine learning approaches often rely on vast amounts of labeled data for every new scenario, a requirement that is not always practical. Although LNNs have shown some capacity to handle out-of-distribution data, a deeper understanding of the principles behind this ability is necessary. Enhancing these principles could enable models to effectively generalize from sparse data and transfer insights gained from past experiences to novel situations. Moreover, combining LNNs with data augmentation strategies holds promise for boosting overall performance, ensuring that the knowledge acquired remains applicable to new challenges without significant degradation. Finally, establishing rigorous evaluation frameworks is essential for accurately measuring the ZSL capabilities of LNNs in real-world wireless communication settings.

## 6.2 Distributed LNNs

Distributing LNNs across various devices and nodes is vital for modern large-scale wireless communication systems. This approach not only boosts scalability, fault resilience, and efficient resource use but also introduces challenges in effective coordination and synchronization. To harness the full potential of distributed LNNs, it is crucial to develop specialized learning algorithms that reduce both communication overhead and latency while implementing robust fault tolerance and dynamic resource management strategies. Federated learning presents an attractive solution by enabling multiple devices to collaboratively train LNNs locally, thereby slashing communication costs and enhancing data privacy. Focusing research on these areas will significantly improve the practical deployment of distributed LNNs in complex wireless environments.

## 6.3 Multi-Modality Fusion

Combining data from multiple modalities (such as sensor data, audio, video, and text) in wireless communication systems can significantly improve the performance and reliability of LNNs. By drawing on these diverse information sources, LNNs develop a richer perspective of the communication environment. However, designing architectures that effectively handle and integrate multi-modal data poses both a challenge and an opportunity. Achieving this goal involves tackling data synchronization and fusion issues across different modalities while ensuring that incorporating multi-modal data enhances overall performance without adding excessive complexity [23].

## 6.4 Training and Inference Latency

A critical area demanding focused future investigation for LNN deployment in 6G is the operational latency, a factor paramount for practical feasibility. Specifically, while LNNs offer unique continuous-time processing, their inference speed must be carefully evaluated. The computational time required to numerically solve the underlying ODEs needs direct measurement and comparison against the stringent, often sub-millisecond, real-time response requirements of demanding 6G applications like ultra-reliable low latency communications (URLLC) or real-time network control. Achieving the necessary inference speeds may require dedicated research into optimized numerical solvers tailored for LNNs, exploring model simplification or approximation techniques, and leveraging hardware acceleration platforms. Equally important, and currently underexplored, is the comprehensive evaluation of the end-to-end training latency. This encompasses not only the time consumed by the training algorithm itself but also the necessary steps of data collection and processing, as well as the subsequent phase of model evaluation. Understanding this complete time cycle is vital, as the highly dynamic nature of 6G environments will likely necessitate frequent model retraining or adaptation to maintain optimal performance. Therefore, future research must dedicate significant effort to quantifying both the inference speed on relevant hardware and the practical duration of the full training pipeline for typical 6G tasks, thereby validating the viability of LNNs within next-generation telecommunication systems.

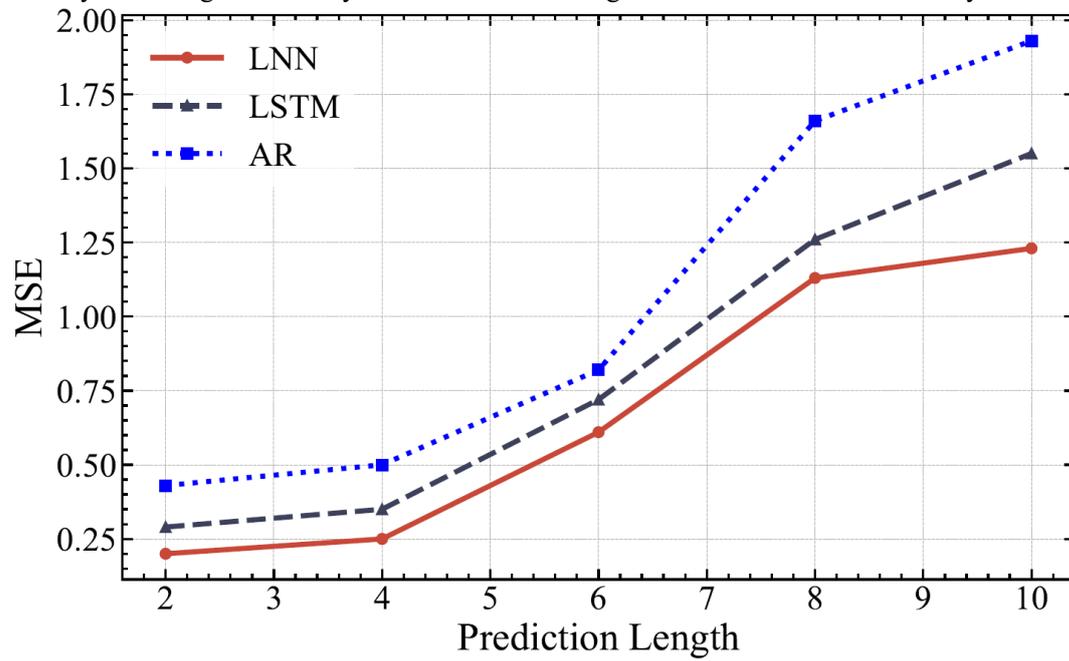

Fig. 4. MSE versus prediction length with real-world CSI.

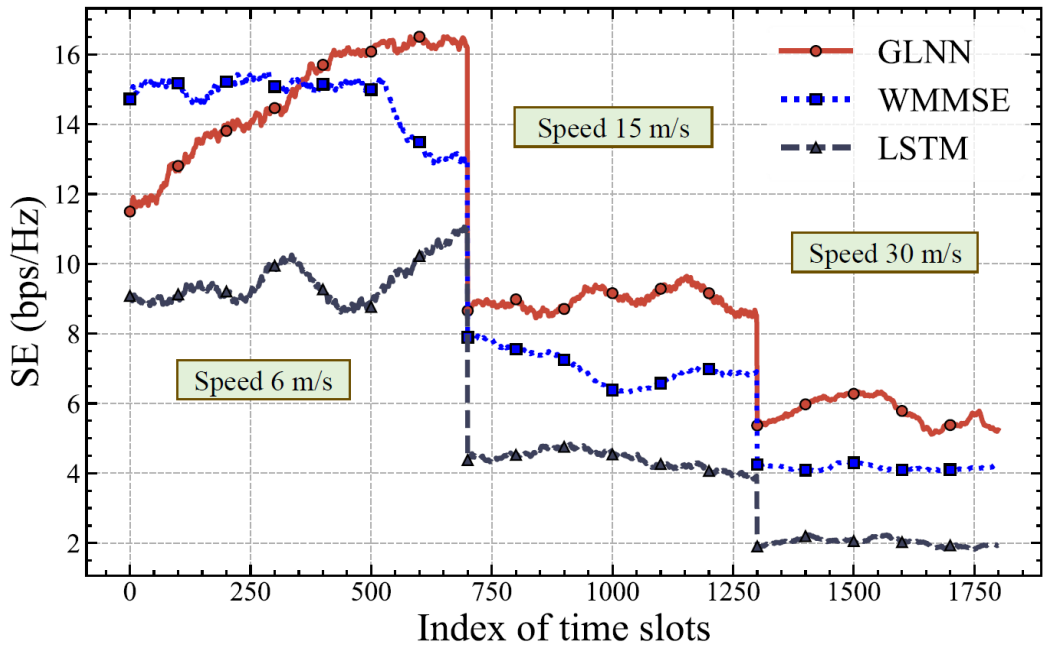

Fig. 5. Average SE in a dynamic beamforming scenario.

Table I
Simulation Parameters in Fig. 4 and Fig. 5

| Parameters | Fig. 4 | Fig. 5 |
| --- | --- | --- |
| BS Antenna Number | 4 | 64 |
| BS Antenna Spacing | $0.5\lambda$ | $0.5\lambda$ |
| User Number | 1 | 4 |
| User Antenna Number | 1 | 2 |
| Central Frequency (GHz) | 6 | 28 |
| Liquid Neuron Number | / | 30 |

## 7. Case Studies

In this section, we summarize the performance of LNNs in two typical application scenarios: channel prediction with LTCs and beamforming with NCPs.

7.1 Channel Prediction with LTCs
We assume an urban microcell scenario where an outdoor base station (BS) serves both outdoor and indoor users [24]. A user is connected to the BS and moves in a random walk with a speed of 2 m/s, with the direction uniformly distributed between 0 and $2\pi$ radians from its initial position. Historical channel state information (CSI) feedback with a length of 20 is utilized to predict future CSI with a length of 5. The test was conducted in a real-world scenario using practical CSI. The field test simulation parameters are summarized in Table I. Fig. 4 illustrates the mean squared error (MSE) versus CSI prediction length in channel prediction. It is evident that the MSE of all schemes increases with prediction length, indicating that longer prediction lengths introduce more uncertainty. Among all schemes, the proposed LTCs-based approach consistently outperforms other baselines, achieving lower MSE, with the performance gap widening as the prediction length increases, particularly when it exceeds 6. This highlights the potential of LTCs in achieving more accurate channel prediction in practical and dynamic scenarios.

7.2 Beamforming with NCPs

We consider a multiple-input multiple-output (MIMO) beamforming system [25]. A BS furnished with M antenna elements concurrently provides service to K users. Each user device possesses Nr antennas. The users experience a range of velocities: 6 m/s, 15 m/s, and culminating in 30 m/s. Each of these phases encompasses 700, 600, and 500 discrete time intervals, respectively. Key simulation settings are enumerated in Table I. The average SE achieved under this dynamic condition, contrasted against alternative benchmark schemes, is illustrated in Fig. 5. The gradient-based liquid neural network (GLNN) approach, leveraging NCPs, rapidly surpasses the WMMSE algorithm after a short initial learning period. It then maintains a superior level of SE when juxtaposed with all other reference systems. This behavior underscores its remarkable capacity for adaptation and its efficacy in environments characterized by temporal variations.

## 8. Conclusion

In this article, we investigated the emerging concept of LNNs that are designed from first principles. We delved into their structure, features, and distinct advantages compared to traditional neural networks, as well as their recent applications. LNNs demonstrate remarkable potential as a key enabling technology in next-generation wireless communications due to their superior generalizability, interpretability, lower complexity, continuous-time modeling capabilities, and robust performance in dynamic environments. By leveraging their adaptive nature and efficient design, LNNs can enhance scalability, fault tolerance, and resource utilization efficiency in wireless networks. However, several challenges remain to be addressed to fully realize the potential of LNNs in practical applications, including improving zero-shot learning capabilities, developing distributed LNN frameworks, integrating multi-modality data, and optimizing cross-layer interactions. Future research is expected to focus on overcoming these challenges to ensure that LNNs can effectively adapt to varying conditions and deliver reliable performance in real-world scenarios. By addressing these issues, LNNs can drive the evolution and enhancement of future wireless networks, paving the way for more adaptive, reliable, and efficient communication systems.

Biographies

**ZHU Fenghao** (zjuzfh@zju.edu.cn) received the B.Eng. degree in information engineering from Zhejiang University, Hangzhou, China, in 2023, and he is currently pursuing the M.S. degree with the College of Information Science and Electronic Engineering, Zhejiang University. His current research interests include massive MIMO, signal processing, and machine learning. Mr. Zhu is a recipient of 2024 IEEE ComSoc Conference Grant.

**WANG Xinquan** (wangxinquan@zju.edu.cn) is currently pursuing the B.Eng. degree at Zhejiang University, Hangzhou, China. His current research interests include 6G, beamforming and machine learning. Mr. Wang is a recipient of 2024 IEEE ComSoc Student Grant.

**ZHU Chen** (zhuc@zju.edu.cn) received his BS degree from North University of China in 2010, and MS degree from Zhejiang University of Technology, China in 2013. He is currently engaged in teaching and research at the College of Engineering, Zhejiang University, China. His main research interests include general sense computing integration, machine learning, image processing, and cloud-edge collaborative computing.

**HUANG Chongwen** (chongwenhuang@zju.edu.cn) obtained his B. Sc. degree in 2010 from Nankai University, and the M. Sc degree from the University of Electronic Science and Technology of China in 2013, and PhD degree from Singapore University of Technology and Design (SUTD) in 2019. From Oct. 2019 to Sep. 2020, he is a Postdoc in SUTD. Since Sep. 2020, he joined into Zhejiang University as a tenure-track young professor. Dr. Huang is the recipient of 2021 IEEE Marconi Prize Paper Award, 2023 IEEE Fred W. Ellersick Prize Paper Award and 2021 IEEE ComSoc Asia-Pacific Outstanding Young Researcher Award. He has served as an Editor of IEEE Communications Letter, Elsevier Signal Processing, EURASIP Journal on Wireless Communications and Networking and Physical Communication since 2021. His main research interests are focused on Holographic MIMO Surface/Reconfigurable Intelligent Surface, B5G/6G Wireless Communications, mmWave/THz Communications, Deep Learning technologies for Wireless communications, etc.